\documentclass[aps,pra,pifont,citesort,epsf,twocolumn,superscriptaddress]{revtex4}
\usepackage{color}
\usepackage{amssymb}
\usepackage{amsthm}
\usepackage[dvips]{graphicx}
\usepackage{float}

\newcommand{\eq}[1]{\begin{equation} \newline #1 \end{equation}}
\newcommand{\eqn}[1]{\begin{eqnarray} \newline #1 \end{eqnarray}}

\newcommand{\ee}{&=&}

\newcommand{\hs}{\hspace{0.2cm}}

\newcommand{\adag}{\hat{a}^{\dag}}
\newcommand{\anih}{\hat{a}}
\newcommand{\x}{\hat{x}}
\newcommand{\p}{\hat{p}}

\newcommand{\bra}[1]{\langle#1|}
\newcommand{\ket}[1]{\left |#1\right \rangle}

\newcommand{\nn}{\nonumber}
\usepackage{relsize}

\newcommand{\matt}[4]{
\left(\begin{array}{cc}
#1 & #2 \\
#3 & #4
\end{array}\right)}

\newcommand{\matfour}[4]{
\left(\begin{array}{cccc}
#1  \\
#2\\
#3 \\
#4 
\end{array}\right)}
\newcommand{\EV}[1]{\left < #1 \right >}
\newcommand{\ga}{g^{\adag\anih}}

\begin{document}

\title[NLA via phase-space transformations]{Non-deterministic noiseless amplification via non-symplectic phase
  space transformations}

\author{Nathan Walk}
\author{Austin P. Lund}
%\email{walk@physics.uq.edu.au}
\author{Timothy C. Ralph}

\address{Centre for Quantum Computation and Communication Technology\\
  School of Mathematics and Physics, University of Queensland, St Lucia,
  Queensland 4072, Australia}

\email{walk@physics.uq.edu.au}
\pacs{03.67.Hk, 42.50.Dv}
\begin{abstract}
  We analyse the action of an ideal noiseless linear amplifier operator,
  $\ga$, using the Wigner function phase space representation. In this setting
  we are able to clarify the gain $g$ for which a physical output is produced
  when this operator is acted upon inputs other than coherent states. We
  derive compact closed form expressions for the action of $N$ local
  amplifiers, with potentially different gains, on arbitrary $N$-mode Gaussian
  states and provide several examples of the utility of this formalism for
  determining important quantities including amplification and the strength
  and purity of the distilled entanglement, and for optimising the use of the
  amplification in quantum information protocols.
\end{abstract}
\maketitle

\section{Introduction}

Optical quantum communication has resulted in numerous protocols that achieve
classically impossible tasks including teleportation
\cite{Bennett:1993p6415,Braunstein:1998p1361}, quantum key distribution
\cite{bb84-orig,Grosshans:2002p377} and super-dense coding
\cite{Bennett:1992p6543,Braunstein:2000p157}.  Furthermore, several of these
have seen experiments ranging in sophistication from proof-of-principle
demonstrations
\cite{Bouwmeester:1997p6498,Furusawa:1998p1451,Bennett:1992p6544,Grosshans:2003p2402,Lance:2005p376}
to implementations approaching real world conditions \cite{Sasaki:5p6496,
Peev:2009p6497,Ma:2012p6413,Yin:2012p5684}. One of the great challenges that
stands between these schemes and the realisation of large scale quantum
information networks is the necessity of preserving often fragile quantum
states in the presence of losses and other decoherence.  A device that allowed
for amplification to combat such effects would be extremely useful, however
the laws of quantum mechanics themselves conspire to enforce a noise penalty
whenever such an operation is attempted \cite{Caves:1982p2405}.

An ingenious recent approach is to circumvent these limits by designing
devices that achieve genuinely noiseless amplifications in a non-deterministic
but heralded manner \cite{proc-disc-2009}.  This noiseless linear amplifier
(NLA) has been the subject of considerable theoretical
\cite{Marek:2010p7535,Fiurasek:2009p7350,Ralph:2011p2764,Walk:2012p5481,Fiurasek:2012p5315,Blandino:2012p5681,Gagatsos:2012p5682,Brask:2012p6414,Ferreyrol:2012p5289,Lee:2011p2177,Kim:2012p7354,NavarreteBenlloch:2012p5680,Gisin:2010p7352,Pitkanen:2011p7353}
and experimental
\cite{Xiang:2010p1449,Ferreyrol:2010p2545,Anonymous:2012p4,Usuga:2010p7394,Ferreyrol:2011p2541,Kocsis:2012p6297,Micuda:2012p5383,Osorio:2012p7349}
work. Applications in quantum key distribution (QKD) with both continuous variable (CV) \cite{Walk:2012p5481,Fiurasek:2012p5315,Blandino:2012p5681} and discrete variables (DV) \cite{Gisin:2010p7352,Pitkanen:2011p7353} have been considered  as well as error correction \cite{Ralph:2011p2764}. In the literature one finds two kinds of analysis.  In the first place one can
consider the ideal amplification operation, which is to implement $\ga$.  In
the amplification regime $(g>1)$ this is an unbounded operator, however for
any particular input state one can always write down a new operator of the
form $\hat{\Pi}_N\ga$ where $\hat{\Pi}_N$ is a projector onto the subspace
spanned by the first $N+1$ energy eigenstates.  This operation will result in
an arbitrarily good approximation of an ideal NLA as $N$ increases at the
price of a decreased, but finite, success probability.  In the second kind of
analysis works thus far have utilised particular linear optics implementations such as that of
the original proposal \cite{proc-disc-2009} or those based upon photon addition and subtraction \cite{Lee:2011p2177,Kim:2012p7354,NavarreteBenlloch:2012p5680}.

Here we will adopt the first approach, and focus on gaining a greater insight
into the properties and applications of the operation $\ga$.  In Section
\ref{t} we will derive the action of the operation on an arbitrary state via
the Moyal product and show that it has the unusual property of being a
Gaussian but non-symplectic map.  Furthermore we address the scenarios in
which, dependent upon the state to be amplified, the NLA fails to transform into a physical output in the limiting procedure
described above. Some
necessary concepts in Gaussian quantum information are introduced in Section
\ref{gp}. In Section \ref{g} we apply our formalism to obtain compact analytic
expressions for up to $N$ amplifiers acting upon N-mode Gaussian states. We give specific examples for one and
two mode cases and comment more rigorously on the physicality of the operation
dependent upon the input state.  In Section \ref{e} we consider in more detail the distribution
of EPR entanglement through a general Gaussian channel which is the situation
most relevant to continuous variable QKD.  We show that
previous attempts to represent the action of the NLA and an effective channel
of the same form but with different parameters are in general insufficient. In a surprising example
we show that in the for large gains the NLA has the effect of transforming an
attack on one half of an EPR pair into an attack upon the other half.  
Finally in Section \ref{c} we conclude.

\section{NLA as a non-symplectic operation \label{t}}

We can define ideal linear amplification in terms of coherent states as
\begin{equation} 
\ket{g\alpha}\bra{g\alpha} =
\Upsilon(\ket{\alpha}\bra{\alpha}).
\end{equation} 
We can realise $\Upsilon$ by
\begin{equation} 
\Upsilon(\rho) = \lim_{N \rightarrow \infty} \Upsilon_N(\rho) 
\end{equation}
where
\begin{equation}
\Upsilon_N(\rho) = p_N^2 g^{a^\dagger a} \Pi_N \rho \Pi_N g^{a^\dagger a}
\end{equation}
where the constant $p_N$ is chosen to make the operation physical and $\Pi_N$ is the projection operator defined earlier.  For most
of this work we will ignore $p_N$, it is important to say a few words about it
at this point.  To achieve ideal linear amplification over the entire harmonic
oscillator Hilbert space requires $\lim_{N\rightarrow \infty}p_N=0$ as the
$g^{a^\dagger a}$ operator has an unbounded spectrum.  However, in any
realistic experiment there will be some bounds within which it is assumed that
the experiment is being performed.  First there will be assumed some energy
bound which can be thought of as a truncation of the Hilbert space.  This
results in accepting a non-unit fidelity with the theoretically ideal
amplifier for states which have a component outside this bound.  The choice of
truncation is somewhat arbitrary, but will generally be determined by the
energy limits of the experiment.  We can think of a sequence of operations
which is indexed by the largest energy eigenstate which is allowed, which we
will call $N$.  For any finite $N$ there is a finite non-zero $p_N$ which one
can choose for $\Upsilon_N$.  As $N$ grows, $p_N$ must reduce.  In the limit
as $N$ approaches infinity, we recover the ideal operation and $p_N$ tends to
zero.  The prediction we make by ignoring $p_N$ in the theory is the state
resulting from this limiting case. However we emphasise that the existing experiments have already shown that for low energy input states approaching the limiting case of ideal operation is achievable without prohibitively low success probability.

The $g^{a^\dagger a}$ operator in Wigner space is 
\eqn{
G_{w}(x,p) \ee\nn
\frac{1}{2\pi}\int e^{ipy}\bra{x-y}\ga\ket{x+y} \hs dy\\ \ee\nn \exp\left\{
\left( \frac{g-1}{g+1} \right) (x^2 + p^2) \right\}\label{Aw},
} 
where we have chosen $\hbar = 2$ and used the identity $\adag\anih =
\frac{1}{4}(\x^2+\p^2-2)$.  Whilst we are interested in the cases where $g>1$, there is no such
restriction needed for the calculations we will perform here.

To act the Wigner representation of the amplifier on an arbitrary input state
requires that the operator product be performed in the Wigner representation.
This is achieved by way of the Moyal Product \cite{Moyal:1949p6600} which we
will denote $\star$.  If we take the input state $\rho$ whose Wigner function
is $W_\rho$ then the Wigner function for the output amplified state
$\rho' = \ga\rho\ga$ is,
\eq{
W_{\rho'} = G_w\star W_\rho\star G_w
}
We will later show that writing the action of the NLA in this form allows the
calculation of compact analytic results for the class of Gaussian states but
already the Wigner representation sheds some light upon the unusual properties
of this operator.

An interesting point noted in the original proposal \cite{proc-disc-2009} is
that the NLA will not produce a physical output when acted upon certain input
states with certain gains.  This does not come as a complete surprise given the state
dependent manner in which the transformation is defined, namely over the
coherent states. The Wigner representation allows us to understand both the
question of physical convergence and Gaussianity of the transformation.

The expression Eq.\ref{Aw} is clearly a Gaussian operation in that it is an
exponential quadratic in the phase space variables however for $g>1$ the
expression is a convex, unbounded function in phase space.  Nonetheless when the
NLA is acted upon a particular input state and the expressions are combined
under the appropriate phase space convolution, Eq.\ref{Aw} tells us that if
the input state is sufficiently `small' (i.e. in terms of the decay of it's
Wigner function in phase space) relative to the gain of the NLA then the
overall output will have a concave, normalisable phase space distribution.
That is, the limiting state $\lim_{N\rightarrow\inf}\Upsilon_N(\rho)$ is a well
defined physical state for this particular situation.  Furthermore in this
case the NLA will preserve the Gaussianity of the input state in the infinite
limit. In section \ref{g} we will make rigourous this argument for the
restricted class of Gaussian states, give exact criteria for the convergence
of the output for a given input.

Finally it is straightforward to see that the operation although mapping
Gaussian input to Gaussian outputs, it is not symplectic.  This is not in
violation of the well known Stone-von~Neumann theorem however as the operation
also fails to be unitary $\ga = (\ga)^\dag \Rightarrow
\ga(\ga)^\dag \neq \mathbb{I}$.  We now turn to the problem of evaluating the
action of the NLA in this phase space representation upon the Gaussian states. First we will introduce some crucial
properties and quantities of these states.

\section{Some Preliminaries of Gaussian Quantum information \label{gp}} 
Within continuous variable quantum information a great deal of attention is
devoted to states with a Gaussian Wigner function and operations that preserve
this form \cite{Weedbrook:2012p5160}.  These so-called Gaussian states and
operations are significant experimentally as they can be efficiently created
and implemented and also lend themselves to an elegant theoretical
description.  This theoretical convenience comes from the fact that although
such states live in in infinite dimensional Hilbert space they can be
completely characterised by their first and second moments.

In particular if we start $N$-mode Gaussian state, $\rho$, living in a tensor
product of $N$ infinite dimensional Hilbert spaces equipped with bosonic
creation and annihilation operators $\anih_1,\adag_1,...\anih_N,\adag_N$ we
can use the corresponding quadrature operators $\x_i = \anih_i + \adag_i$ and
$\p_i = i(\adag_i - \anih_i)$ as phase space coordinates and completely
characterise the state as follows: grouping the quadratures together in a
vector ${\bf r} := (\x_1,\p_1,...,\x_N,\p_N)$ we define any Gaussian state by
a displacement (mean) vector, 
\eqn{
{\bf d} = \mathrm{tr}(\rho {\bf r})
}
and covariance matrix
\eq{
{\bf \Sigma}_{ij} = 
 \mathrm{tr} \left( 
   \rho \{({\bf r}_i - {\bf d}_i),({\bf r}_j - {\bf d}_j)\}_+
 \right)
}
where $\{\}_+$ is the anti-commutator. Matrices will be denoted by boldface throughout
and products of such terms should be interpreted as matrix multiplication.
 For a given square matrix to be a
legitimate covariance matrix (CM) it must satisfy the uncertainty principle
which in this formalism is the following positive semi-definiteness condition
\cite{Simon:1994p6914},
\eq{
  {\bf \Sigma} + i{\bf \Omega} \geq 0
}
where
\begin{equation} 
{\bf \Omega} = \left(
  \begin{array}{cc} 
    0 & 1 \\ 
    -1 & 0
  \end{array}
\right)
\end{equation}
which is called the symplectic form.

The purity of a Gaussian state is obtained simply via the determinant of the
CM 
\eq{
\mu := \mathrm{tr}(\rho_G^2) = \frac{1}{\sqrt{\det ({\bf \Sigma})}}
\label{purity}
}
Thus a Gaussian state is pure if and only if we have $\det({\bf \Sigma}) = 1$.

Likewise the entanglement of a Gaussian state is completely determined by the
second moments, in particular their symplectic spectra which we obtain by
making use of Williamson's theorem \cite{Williamson:1936p7102}.  For any
N-mode CM ${\bf \Sigma}$ there exists a symplectic diagonalisation given by
\eq{
{\bf \Sigma = SWS^T}, \hs {\bf W} = \bigoplus_{k=1}^N \lambda_k \mathbb{I},
}
where ${\bf S}$ is a symplectic matrix and $\lambda_k\geq1$ are the N
symplectic eigenvalues of ${\bf \Sigma}$.  They can be computed either by
solving an $N^{th}$ order polynomial where the coefficients are so-called
symplectic invariants \cite{Serafini:2006p5052} or by finding the standard
eigenspectrum of $|i{\bf \Omega \Sigma}|$ where the absolute value is used in
the operatorial sense \cite{Weedbrook:2012p5160}.  Using the fact that for
symplectic matrices $\det({\bf S})=1$ the physicality condition for a CM can
be rewritten as,
\eq{
{\bf \Sigma}>0,\hs \lambda_-\geq 1 \label{lambda}
} 
where $\lambda_-$ is the smallest symplectic eigenvalue

A common bipartite entanglement criteria is to consider the partial transpose
(PT) of the density matrix with respect to one subsystem
\cite{Peres:1996p7116,Horodecki:1996p7117}, with entanglement (separability)
corresponding to the non-physicality (physicality) of the resultant operator.
For Gaussian states a similar results hold based upon the CM condition, namely
that if we consider the partial transpose of a bipartite $N\times M$ mode CM
\eq{{\bf \tilde{\Sigma}} = (\mathbb{I}_A \oplus {\bf T}_B) {\bf \Sigma}
(\mathbb{I}_A \oplus {\bf T}_B)} where \eq{{\bf T}_B := \oplus_{m=1}^M
\matt{1}{0}{0}{-1} } then entanglement corresponds to the non-physicality of
${\bf \tilde{\Sigma}}$. It is straightforward to show that we always have
${\bf \tilde{\Sigma}>0} $therefore by Eq.\ref{lambda} the criterion boils down
to checking \eq{\tilde{\lambda}_- \geq1.\label{sep}}

It should be noted that while Eq.\ref{sep} is generally a necessary condition
for bipartite separability it is only necessary and sufficient for $1\times N$
\cite{Werner:2001p7157} and the bisymmetric class of $N\times M$ Gaussian
states \cite{Adesso:2012p5314}.  For such states however one can quantify the
entanglement by means of the log negativity \cite{Vidal:2002p7156} which
can be computed for continuous variables \cite{Simon:2000p6698} from the
smallest PT symplectic eigenvalue via, \eq{\mathcal{E} = \max \{ 0, \log
\tilde{\lambda}_-\}\label{logneg}}

Finally a large class of relevant operations and channels (squeezing, passive
mode mixing, amplification and white noise channels) fall in the class of
Gaussian operations.  These can then be compactly implemented at the level of
covariance matrix transformations.  In this formalism, Gaussian unitary
operations correspond to transformations of the form ${\bf
\Sigma}_{\mathrm{out}} = {\bf S}{\bf \Sigma}_{\mathrm{in}}{\bf S}^T$ where
${\bf S}$ is a symplectic matrix.  The NLA does not take this symplectic form,
nonetheless we will derive analytic input output relations for the
amplification of arbitrary Gaussian states.

\section{Amplification of Gaussian states\label{g}}

It is inconvenient to evaluate the $\star$ operation in the $x$ and $p$
variables.  It is more useful here to write this product down in terms of the
characteristic functions $(\chi)$ which are the Fourier transform of the
corresponding Wigner functions.  Here we choose those variables to be $a$ and
$b$ corresponding to transformed $x$ and $p$ respectively.  With our choice of
$\hbar = 2$ (such that the vacuum noise is unity) the product in the Fourier
transformed space for operators $A$ and $B$ is:
\begin{equation} 
\chi_{AB} (a,b) = \frac{1}{(2\pi)^2} \int {da^\prime}{db^\prime} 
\chi_A(a-a^\prime,b-b^\prime) e^{i \mathbf{a}^T {\bf \Omega} \mathbf{a'}}
\chi_B(a^\prime,b^\prime)
\end{equation} 
where $\mathbf{a} = (a,b)^T$, $\mathbf{a'} = (a',b')^T$.  This formalism
carries over to the multidimensional case fairly naturally with the matrix
${\bf \Omega}$ extended using the direct sum.  The characteristic function for
multidimensional Gaussians is
\begin{equation} 
\chi_G = e^{i\mathbf{{\bf d}}^T\mathbf{a}} 
e^{-\frac{1}{2} \mathbf{a}^T {\bf \Sigma} \mathbf{a}}
\end{equation}
The Moyal product of two Gaussian will then result in another Gaussian.  Thus we can simply read off the output
covariance matrix and hence overall state from the output characteristic
function. Also, we can consider acting amplifiers on all modes and the
amplification transformation extends naturally from Eq.\ref{Aw} where we rewrite as 
\eq{G_w(x,p) = \exp\left\{ {\bf
G}^{-1} (x^2 + p^2) \right\}}  we have defined the matrix 
\eq{
{\bf G}=\matt{\frac{g+1}{g-1}}{0}{0}{\frac{g+1}{g-1}}
} 
and extend
${\bf G}$ to the multimode case by taking the direct sum of all the individual modes.  We can take the
limit as $g \rightarrow 1$ for modes which have no amplification. This limit
constitutes a delta function as expected.

To conjugate the density operator by the action of an ideal amplifier the
Moyal product must be applied twice.  For a Gaussian density operator with
mean vector ${\bf d}$ and covariance matrix ${\bf \Sigma}$ this computation
gives two equations for the mean vector and covariance matrix, one from each
application of the product.

To compute the relationships for the mean vector and covariance matrix we will
utilize the result of the gaussian integral
\begin{equation}
\int dx_1 dx_2 \cdots dx_n e^{-\frac{1}{2}{\bf x^T \cdot {\bf A} \cdot x} + {\bf {\bf b} \cdot x}}
\propto e^{\frac{1}{2} {\bf {\bf b}^T \cdot {\bf A}^{-1} {\bf b}}}
\end{equation}
where this equation holds for matrix ${\bf A}$ and vector ${\bf b}$ possibly complex (note
that the transpose is taken and not the adjoint).  We can ignore the
proportionality factor here as it can be considered part of the $p_N$ terms we defined in Section\ref{t}. Since we are only interested in the convergent state this factor
is not relevant.

We will start by computing the left product by the $\ga$ operator $\rho_1 =
\ga \rho$ (which is not a physical state) using the Wigner representation
and the Moyal product.  Substituting in the definitions for the operators
results in the integral (ignoring constant proportionality factors)
\begin{equation}
W_{\rho_1} = \int d\mathbf{a}^\prime e^{(\mathbf{a} - \mathbf{a}^\prime)^T
{\bf G} (\mathbf{a} - \mathbf{a}^\prime) + i \mathbf{a}^{\prime T} {\bf {\bf
\Omega}} \mathbf{a} + i {\bf {\bf d}}^T \mathbf{a}^\prime - \frac{1}{2}
\mathbf{a}^{\prime T} {\bf {\bf \Sigma}} \mathbf{a}^\prime}
\end{equation}
We can now rearrange the polynomial to obtain an expressions more like that of
the standard Gaussian integral.
\begin{equation}
W_{\rho_1} = e^{\frac{1}{2} \mathbf{a}^T {\bf G} \mathbf{a}} \int
d\mathbf{a}^\prime e^{(i({\bf {\bf d}}^T + \mathbf{a}^T{\bf {\bf \Omega}}) -
\mathbf{a}^T {\bf G})\mathbf{a}^{\prime} -\frac{1}{2} \mathbf{a}^{\prime T} ({\bf
{\bf \Sigma}} - {\bf G}) \mathbf{a}^\prime}
\end{equation} 
Upon evaluating this integral we find that the covariance matrix transforms as
\begin{equation}
{\bf {\bf \Sigma}}_1 = (i {\bf {\bf \Omega}} - G) ({\bf {\bf
\Sigma}} - G)^{-1} (i {\bf {\bf \Omega}} + G) - G
\end{equation} 
and the mean vector transforms as s
\begin{equation} {\bf d}_1 =
(i {\bf \Omega} - G) ({\bf \Sigma} - G)^{-1} {\bf d}
\end{equation}

Following a similar calculation for right multiplying by the amplification
operator transforms the covariance matrix as
\begin{equation} 
{\bf \Sigma}_{out} = (i {\bf \Omega} + {\bf \Sigma}_1) ({\bf
\Sigma}_1 - {\bf G})^{-1} (i {\bf \Omega} + {\bf \Sigma}_1) + {\bf \Sigma}_1
\end{equation}
and the mean vector transforms as
\begin{equation}
{\bf d}_{out} = {\bf d}_1 - (i {\bf \Omega} + {\bf \Sigma}_1)
({\bf \Sigma}_1 - {\bf G})^{-1} {\bf d}_1.
\end{equation}
In the single mode case, ${\bf G}$ will be proportional to the identity and
hence this can be used to greatly simplify the expressions, but we will work
with the case of ${\bf G}$ not being proportional to the identity and hence
will include multimode cases where different gain variables may be used on
different modes.

To simplify these expressions it is important to note that
\begin{equation} 
(i{\bf \Omega} - {\bf G})(i{\bf \Omega} + {\bf G}) =
\mathbb{I} - {\bf G}^2.
\end{equation}
This holds as ${\bf G}$ is diagonal and for each local block the two entries
are the same.  This gives ${\bf \Omega} {\bf G} = {\bf G} {\bf \Omega}$ and
hence the above relation.

We will now dissect the terms in ${\bf \Sigma}_{out}$ and evaluate them:
\begin{eqnarray}
({\bf \Sigma}_1 - {\bf G})^{-1} & = & (i{\bf \Omega} + {\bf G})^{-1}\nn \\
&&\left[
({\bf \Sigma} - {\bf G})^{-1} + (i{\bf \Omega}+{\bf G})^{-1} - (i{\bf \Omega}-{\bf G})^{-1}
\right]^{-1} \nn \\
&&(i{\bf \Omega} - {\bf G})^{-1} 
\end{eqnarray}
\eqn{i{\bf \Omega} + {\bf \Sigma}_1 & = & (i{\bf \Omega} - {\bf G})\nn \left[ ({\bf \Sigma} - {\bf G})^{-1} + (i{\bf \Omega}+{\bf G})^{-1} \right]\\
&&(i{\bf \Omega} + {\bf G}) }
\eqn{i{\bf \Omega} - {\bf \Sigma}_1 & = & (i{\bf \Omega} - {\bf G})\left[ -({\bf \Sigma} - {\bf G})^{-1} + (i{\bf \Omega}-{\bf G})^{-1} \right]\nn\\
&& (i{\bf \Omega} + {\bf G})}

Substituting these results back gives
\begin{equation}
{\bf \Sigma}_{out} = \left[ ({\bf \Sigma}-{\bf G})^{-1}-2({\bf G}^{-1}-{\bf G})^{-1} \right]^{-1} - {\bf G}
\end{equation}
This can be further rearranged to give
\eqn{
{\bf \Sigma}_{out} \ee {\bf G}^{-1}\left({\bf G}^{-1} + {\bf G} - 2{\bf \Sigma} \right)^{-1}({\bf \Sigma}-{\bf G})\nn\\
&+&{\bf G}\left({\bf G}^{-1} + {\bf G} - 2{\bf \Sigma} \right)^{-1}({\bf \Sigma}-{\bf G}^{-1})
}
If instead of using the matrix ${\bf G}$ with diagonal elements of the form
$\frac{g+1}{g-1}$, we use a matrix 
\eqn{
{\bf g} = \matt{g}{0}{0}{g}
} 
extended to many modes also using the direct sum, then this matrix equation
becomes
\begin{eqnarray}
{\bf \Sigma}_{out} \ee {\bf g} \left[ {\bf g}^2 + 1 - {\bf \Sigma} ({\bf g}^2 - 1) \right]^{-1}\nn\\
&&\left[ {\bf \Sigma} ({\bf g}^2 + 1) - ({\bf g}^2 - 1) \right] {\bf g}^{-1}\label{geq}
\end{eqnarray}
This equation can be further simplified if we replace the matrix of linear
gains ${\bf g}$ to a matrix of logarithmic gains 
\eq{
{\bf l} =\matt{\ln g}{0}{0}{\ln g}
}
\begin{equation}
{\bf \Sigma}_{out} = (\cosh {\bf l} - {\bf \Sigma} \sinh {\bf l})^{-1} ({\bf \Sigma} \cosh {\bf l} - \sinh {\bf l}).
\end{equation}

Using the same relationships found above, we can now substitute into the
expression for ${\bf d}_{out}$ and get
\begin{equation}
{\bf d}_{out} = \left[ 2({\bf \Sigma} - {\bf G})({\bf G} - {\bf G}^{-1})^{-1} - 1 \right]^{-1} {\bf d}
\end{equation}
and using the ${\bf g}$ matrix form gives
\begin{equation}
{\bf d}_{out} = 2{\bf g}\left[ {\bf g}^2 + 1 - {\bf \Sigma} ({\bf g}^2 - 1) \right]^{-1} {\bf d}\label{deq}
\end{equation}
and using the logarithmic gain form gives
\begin{equation}
{\bf d}_{out} = ( \cosh {\bf l} - {\bf \Sigma} \sinh {\bf l})^{-1} {\bf d}
\end{equation}

We will now use the results from this method to illuminate how we can easily
compute the action of the NLA on Gaussian states in one and two-modes.

\subsection{Single-mode states}
In general one can unitarily transform one mode Gaussian to remove cross-correlations between the
quadratures and hence write CM's of diagonal form,
\eq{\nn {\bf \Sigma} = \matt{V_x}{0}{0}{V_p}}
with relevant examples including thermal, squeezed and coherent states.
Considering such states with input mean vector ${\bf d} = (\EV{\x},\EV{\p})^T$
we transform via \ref{deq} to find
\eqn{{\bf d}_{NLA} = 
\left(
\begin{array}{c}
\frac{2 g\EV{x} }{V_x+1-g^2 (V_x-1)}\\
\frac{2 g\EV{p} }{V_p+1-g^2 (V_p-1)} 
\end{array}
\right)
}
\begin{figure}[htbp]
\centering
\includegraphics[width = 8.3cm]{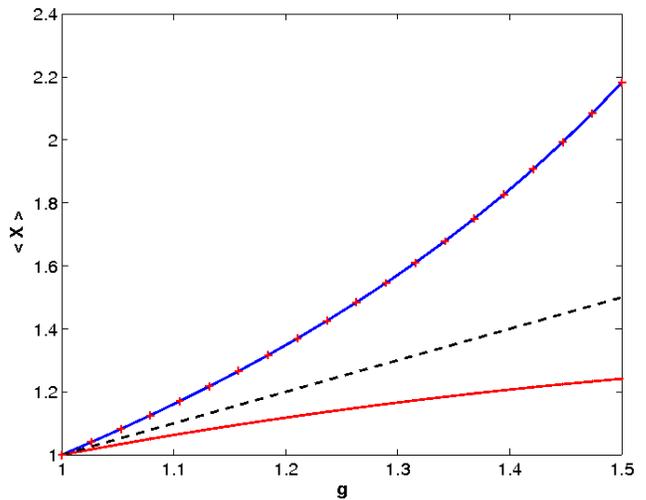}
\caption{\label{dout} (colour online) Expectation values of output quadratures
  $X = \{\x,\p\}$ for input coherent (black, dashed), thermal (blue, solid)
  and squeezed (red, $\x$ dot-dash, $\p$ crosses) states. For all plots V =
  1.5 and ${\bf d} = (1,1)$.}
\end{figure}

Naturally for coherent states ($V_x = V_p = 1$) the above expression reduces
to ${\bf d}_{\mathrm{out}} = g{\bf d}$ but for thermal states ($V_x = V_p =
V>1$) the amplification is no longer linear in $g$. Instead it increases
rapidly and becomes infinite at a maximum value of the gain before becoming
negative and as we will see unphysical.

While both thermal and coherent states are phase symmetric, this is not true
of squeezed states ($V_x = V^{-1}, V_p = V>1$) and thus the two quadrature
displacements amplify differently. While the anti-squeezed displacement
transforms identically to the thermal case the squeezed quadrature
displacement amplifies sub-linearly in $g$.  These results are plotted in
Fig.\ref{dout}

\begin{figure}[htbp]
\centering
\includegraphics[width = 8.3cm]{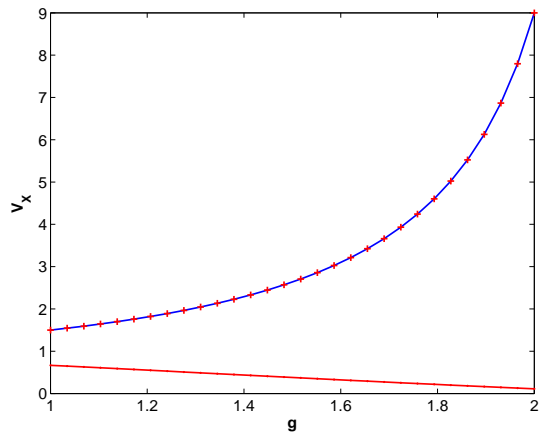}
\caption{\label{Vout1m} (colour online) Quadrature variances $V_X$, $X =
  \{\x,\p\}$ for input thermal (blue, solid) and squeezed (red, $\x$ dot-dash,
  $\p$ crosses) states. For all plots V = 1.5 and ${\bf d} = (1,1)$.}
\end{figure}

Turning to the output covariance matrix we can substitute int Eq.\ref{geq} to
obtain,
\eqn{{\bf \Sigma}_{NLA} = 
\left(
\begin{array}{cc}
 \frac{V_x+1+g^2 (V_x-1)}{V_x+1-g^2 (V_x-1)} & 0 \\
 0 &  \frac{V_p+1+g^2 (V_p-1)}{V_p+1-g^2 (V_p-1)} 
\end{array}
\right).
\label{1mcm}}
Following a similar pattern the thermal state variance and that of the
anti-squeezed quadrature for the squeezed state become infinite at the same
gain for which the squeezed variance vanishes.  In other words a thermal state
thermalises further whereas a squeezed state is squeezed further
Fig.\ref{Vout1m}.  Note that this is true regardless of the angle of
squeezing. This phase insensitive squeezing property was addressed in
\cite{Gagatsos:2012p5682} where the potential causal paradoxes were resolved.

This behaviour is in accord with our intuition from the first section in which
we saw evidence for a critical gain at which the output state was described by
a flat phase space distribution (infinite variance) and beyond which was
unphysical (negative variance).

By considering the output on the single mode case we can derive the necessary
and sufficient condition for convergence to a physical output state. Solving
for the singularity in the output CM Eq.\ref{1mcm} we find that for a single
NLA applied to one mode of a multi-mode Gaussian state the maximum physical
gain is
\eq{g_{\mathrm{max}} = \frac{V+1}{V-1}.}
where $V$ is the variance of the input to the amplifier.  Note that in the multi-mode case
where more than one NLA is present this condition will be necessary but no
longer sufficient to guarantee a convergent output.

\subsection{Two-mode states}

The canonical bipartite Gaussian state is the two-mode squeezed vacuum or EPR
state which, among other applications, is the building block for quantum
teleportation and the theoretical analysis of continuous variable quantum key
distribution.  In the number basis it has the form,
\eq{\ket{EPR} = \sqrt{1-\chi^2}\sum_n \chi^n\ket{n}\ket{n}} 
where $\chi$ ranges from 0 for an unsqueezed vacuum up to unity for an
infinitely squeezed, maximally entangled state.

The effect of the NLA in this context has already been considered in
\cite{Ralph:2011p2764} where distillation in the presence of loss was
demonstrated and it has also been shown to benefit key distribution over
general Gaussian channels
\cite{Blandino:2012p5681,Walk:2012p5481,Fiurasek:2012p5315}.  Here we will
analytically re-derive the previous results but with much less effort via our
new formalism.

An arbitrary Gaussian channel can be described by a transmission $T$ and a
thermal noise parameter, for instance the variance of the enivronment $V_E$.

The covariance matrix of an EPR state with mode $B$ distributed through such a
channel is given by,
\eqn{{\bf \Sigma}_{\mathrm{in}}  = \matt{V_A\hspace{2mm}\mathbb{I}_2}{c_{AB}\hspace{2mm}\sigma_z}{c_{AB}\hspace{2mm}\sigma_z}{V_B\hspace{2mm} \mathbb{I}_2}\label{cm}}
with $\sigma_z = [1,0;0,-1]$ and
\begin{eqnarray}
V_A &=& \frac{1+\chi^2}{1-\chi^2}\nonumber\\
V_B &=& T\frac{1+\chi^2}{1-\chi^2} + (1 - T)V_E \nonumber\\
c_{AB} &=&  \frac{2\sqrt{T}\chi}{\chi^2-1}\label{abc}
\end{eqnarray}
Again applying Eq.\ref{geq} where an NLA of gain $g$ is applied to mode $B$ we
find an output covariance matrix of the same form as Eq.\ref{cm} but with
entries related to the inputs via
\begin{eqnarray}
V_A' &=& \frac{1}{N}\hs\left [ V_A (V_B+1)- T (V_A^2-1) \right . \nn\\
 &+&\left.  g^2 (T (-1 + V_A^2) - V_A (-1 + V_B))\right]\nonumber\\
V_B' &=&\frac{1}{N}\hs\left[ V_B+ 1 + g^2 (V_B-1) \right] \nonumber\\
c_{AB}' &=& \frac{1}{N}\hs 2 g \sqrt{T (V_A^2 -1)}\label{abcnla}
\end{eqnarray}
where 
\eq{N = V_B+1 - g^2 (V_B-1)}
which must be positive and non-zero and hence gives the constraint on $g$
outlined in the previous section.

\begin{figure}[htbp]
\centering
\includegraphics[width = 8.5cm]{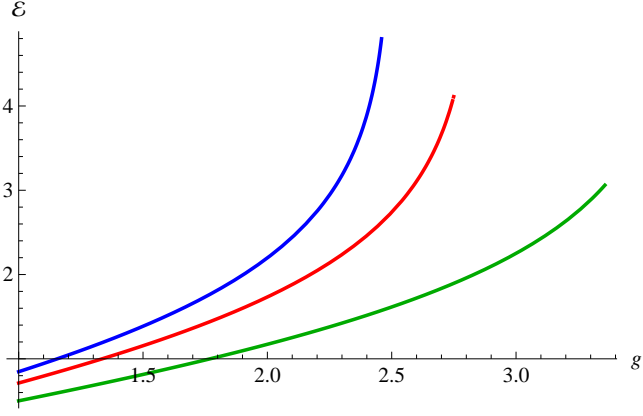}
\caption{\label{logfig} (colour online) Logarithmic negativity of an amplified
  EPR state as a function of gain for varying levels of channel
  decoherence. We paramaterise this by fixing an initial EPR strength of $\chi
  = 0.4$ $(V_A\approx1.4)$ and a thermal environment of variance $V_E = 1.1$
  and mixing the two on beamsplitter of decreasing transmission $T$. Plotted
  curves are for $T=1$ (blue), $T=0.8$ (red) and $T=0.5$.}
\end{figure}

We can now analyse the output states in terms of their entanglement and
purity. We know from\cite{proc-disc-2009} that for an initially pure EPR
state, as the NLA gain reaches it's maximum value the resultant output will
tend towards a pure maximally entangled EPR state. The same results were also shown for pure loss channels where, crucially, no entanglement is generated between the transmitted mode and the environment.
As soon as there is
decoherence any amount of decoherence the NLA wtill start to distill
correlations between the amplified mode and the noisy environment as well as
the other arm of the EPR.  Thus it becomes impossible to distill maximal
entanglement within the allowable gain range. We will consider the
entanglement distillation in the presence of Gaussian decoherence via the
logarithmic negativity given in Eq.\ref{logneg}. For two-mode states with CM
of the form Eq.\ref{cm} the symplectic eigenvalues of the PT state are given
by,
\eq{\tilde{\lambda}_{\pm} = \sqrt{\frac{\Delta \pm \sqrt{\Delta^2 - 4 \det({\bf \tilde{\Sigma}})}}{2}}}  
where $\Delta:= a^2 + b^2 + 2c^2$ and $\det({\bf \tilde{\Sigma}}) = \det({\bf
  \Sigma}) = (ab - c^2)^2$.  In Fig.\ref{logfig} we plot the logarithmic
negativity as a function of gain for the ideal channel and for increasing
levels of decoherence.

We paramaterise the channel by fixing the initial variance ($V_A$) (or
alternatively the EPR paramater $\chi$) of the EPR to be distributed and the
variance of a thermal environment mode $V_E$ and then interacting the two
modes on a beamsplitter of varying transmission. A perfect channel corresponds
to $T=1$ with an increasingly mixed output for smaller values. For the cases
plotted we have $V_E<V_A$ so as the transmission decreases so does Bob's
variance leading to a higher maximum allowable gain.  These results show the
entanglement increasing monotonically as a function of $g$. While a perfect
channel allows for maximum distillation, modest amounts of loss and noise
swiftly decrease the maximum distillable entanglement and the rate of increase
with $g$.

We are also interested in the purity of our final state, and by considering
Eq.\ref{purity} it is swiftly apparent that the purity decreases monotonically
with gain and that
\eq{\lim_{g\rightarrow gmax} \mu = 0.}
We plot the purity of the output states corresponding to those shown in
Fig.\ref{logfig} demonstrating the degradation of purity as a function of
gain. We see that the decay is initially gentle and then rapidly decreases as
the maximum gain is approached. Thus we are left with a further
restriction upon the maximum distillable entanglement if we would also like to
maintain high levels of purity.
\begin{figure}[htbp]
\centering
\includegraphics[width = 8.5cm]{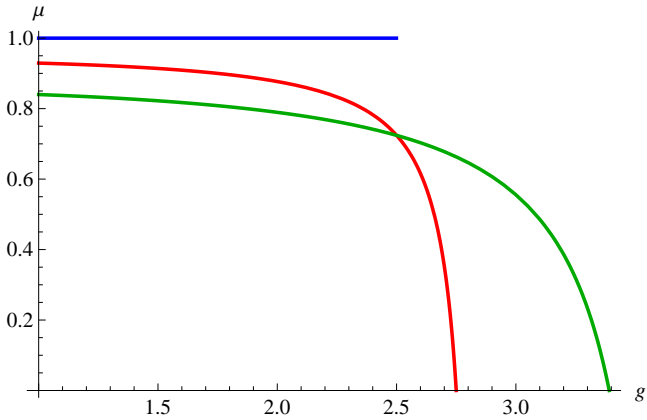}
\caption{\label{purityfig} (colour online) Purity of an amplified EPR state as
  a function of gain for the same input parameters as Fig.\ref{logfig}. Once again Plotted
  curves are for $T=1$ (blue), $T=0.8$ (red) and $T=0.5$}
\end{figure}

These results demonstrate the competing considerations of entanglement
strength and purity that must be taken into account when choosing the NLA gain
to be applied. In the previous discussion ignored the final degree of freedom available
to Alice and Bob, namely the strength of the input EPR. For example in a
situation where the purity of entanglement is of great importance Alice and
Bob can attempt to improve their protocol by starting with weaker entanglement
and amplifying it further after the channel. As a further example of the
utility of our formalism is the ease with which one can optimise over input
parameters to maximise a desired operational quantity. As a demonstration we
will plot the maximum achievable fidelity between the de-cohered and
subsequently amplified EPR state and a pure target EPR state of a certain
strength. We plot the results as a function of the target EPR parameter
$\chi_T$ where both input EPR strength and the NLA gain have been
optimised over. The fidelity between two-mode Gaussian states of zero mean is
given by \cite{Marian:2012p6359},
\eq{F = \frac{1}{\sqrt{\Gamma} + \sqrt{\Lambda} - \sqrt{(\sqrt{\Gamma} + \sqrt{\Lambda} )^2- \Theta} }}
where 
\eqn{\Gamma \ee\nn\frac{1}{16} \left[1-2 c_{AB} c_{AB}'+V_A \left(-V_B \left(c_{AB}'\right){}^2+V_A'\right)\nn \right . \\
 &+&\left. V_B \left(1+V_A V_A'\right) V_B'+c_{AB}^2 \left(\left(c_{AB}'\right){}^2-V_A' V_B'\right)\right]{}^2\nn\\
\Lambda \ee\nn \frac{1}{16} \left(c_{AB}^4+c_{AB}^2 \left(2-2 V_A V_B\right)+\left(-1+V_A^2\right) \left(-1+V_B^2\right)\right) \nn\\
&\times& \left[\left(c_{AB}'\right){}^4+\left(c_{AB}'\right){}^2 \left(2-2 V_A' V_B'\right)+\left(-1+\left(V_A'\right){}^2\right)\right.\nn\\
&&\left. \left(-1+\left(V_B'\right){}^2\right)\right]\nn\\
\Theta \ee \frac{1}{16} \left(\left(c_{AB}+c_{AB}'\right){}^2-\left(V_A+V_A'\right) \left(V_B+V_B'\right)\right){}^2}

\begin{figure}[htbp]
\centering
\includegraphics[width = 8.5cm]{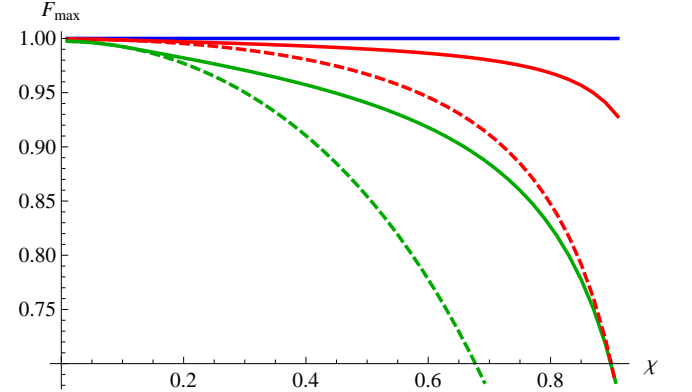}
\caption{\label{fidopt} (colour online) Maximum fidelity between a distributed
  EPR state after amplification and a pure target EPR state as a
  function of the target EPR strength $\chi_T$ for varying levels of channel
  decoherence. We paramaterise the channel by fixing a thermal environment of
  variance $V_E = 1.01$ and mixing the two on beamsplitter of decreasing
  transmission $T$. Plotted curves are for $T=1$ (blue), $T=0.9$ (green) and
  $T=0.5$ red). For all points the NLA gain and input EPR strength are
  simultaneously optimised over. To illustrate the improvement due to amplfication we also plot the performance in the absence of an NLA (dashed lines) for comparison.}
\end{figure}
where the input covariance matrices are in standard form with entries $V_A,$
$V_B$, $c_{AB}$ and $V_A',$ $V_B'$, $c_{AB}'$ respectively. This expression differs slightly from that given in \cite{Marian:2012p6359} as we have a different noise convention with our vacuum normalised to 1 instead $\frac{1}{2}$.

\section{Effective circuit\label{e}}
In the previous section we derived the effect of the NLA upon an EPR state
transmitted through an arbitrary Gaussian channel and calculated the strength
and purity of the resultant entanglement.  The same situation was also
considered in \cite{Blandino:2012p5681,Walk:2012p5481,Fiurasek:2012p5315} for
the purposes of Continuous Variable Quantum Key Distribution (CVQKD) where the
authors paramaterised the results in terms of an effective combination of a
different EPR state and channel.  However if we pursue this representation in
detail we find that although such a description is helpful as a tool for
conceptualising how the NLA will affect communication rates it is insufficient
to describe the true nature of correlations with an eavesdropper (Eve).

One can attempt to solve for a set of effective parameters $T',\xi',\chi'$,
where the $\xi$ is another noise parameter more commonly used the CVQKD
literature. It is defined by setting the total variance added by the channel
environment to be $V_E = \frac{1-T+T\xi}{1-T}$. From this formula it is
apparent that the added channel noise has been split up into a component due
to loss and the so-called excess noise $\xi$. In these papers the output of
the NLA corresponds to a scenario where Alice created a stronger EPR initially
and transmitted it through an effective channel. Setting the equations in
Eqs.\ref{abc} and \ref{abcnla} equal we obtain,
\eqn{\chi' \ee\nn \sqrt{1+\frac{2T(1-g^2))}{g^2 T \xi -T \xi -2)}}\chi\\
T' \ee\nn \frac{4 g^2 T}{(-2 + (-1 + g^2) T (-2 + \xi)) (-2 + (-1 + g^2) T \xi)}\\
 \xi' \ee -\frac{1}{2} (-2+(-1+g^2) T (-2+\xi )) \xi. \label{param}}  
The secret key rates of the previous works are of course correct as they only
depend upon the reduced covariance matrix shared by Alice and Bob, however
this interpretation is not always valid and it turns out does not fully
capture the correlations generated by the NLA.

In general one assumes that all of the observed noise originates from the
eavesdropper interactions, in other words that Eve purifies the state.  For a
noisy Gaussian channel where Bob's mode is effectively mixed with a thermal
state a valid purification is for Eve to have her own EPR state and interact
one arm with Bob's mode.

Consider the symmetrical case where Alice and Eve create identical EPR pairs
and interact them upon a 50:50 beamsplitter.  If Bob applies an NLA and we ask
about his correlations with Alice and then Eve then the previous analysis
results in a contradiction. If we consider the Alice-Bob channel we we see the
effective transmission increase, but if we consider the Eve-Bob channel
exactly the same result should hold.  Clearly the beamsplitter ratio cannot
simultaneously increase and decrease and we arrive at a contradiction.

\begin{figure}[htbp]
\centering
\includegraphics[width = 8.5cm]{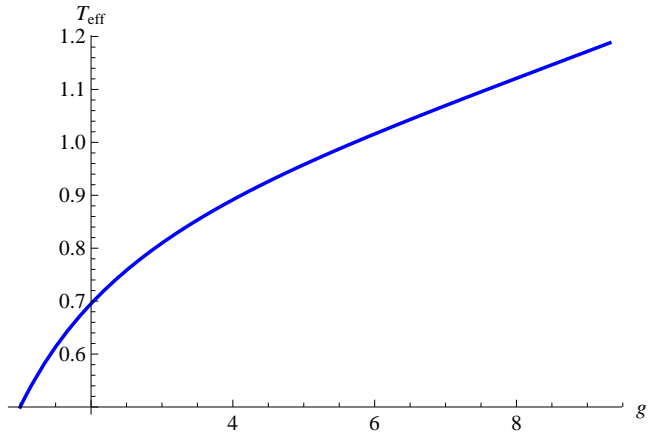}
\caption{\label{teffold}
Effective channel transmission as a function of gain, as given by Eq.\ref{param}. Input parameters are the same as Fig.\ref{logfig} with $T=0.5$. The curve is only plotted for gains less than the maximum allowed value but the effective transmission still exceeds the maximum sensible value for a beamsplitter.}
\end{figure}

In fact if we plot the effective transmission given in Eq.\ref{param} as a
function of gain, Fig.\ref{teffold}, we see that for large but allowable gains
they cease to make physical sense.  In particular the effective channel
transmission can surpass unity, indicating that the NLA can not always
effectively be equated with a beamsplitter interaction.  Note that although
this effective paramaterisation has broken down, key rates calculated by
considering only the reduced covariance matrix are still valid as the
entropies are independent of Eve's particular purification.  Nonetheless we
are interested in ascertaining the exact form of the interaction given by the
NLA.

The ease with which our method can be adapted to a multi-mode picture allows
us to straightforwardly answer this question by explicitly including Eve's
modes in the calculation and explicitly analysing the correlations.
\begin{figure}[htbp]
\centering
\includegraphics[width = 8.5cm]{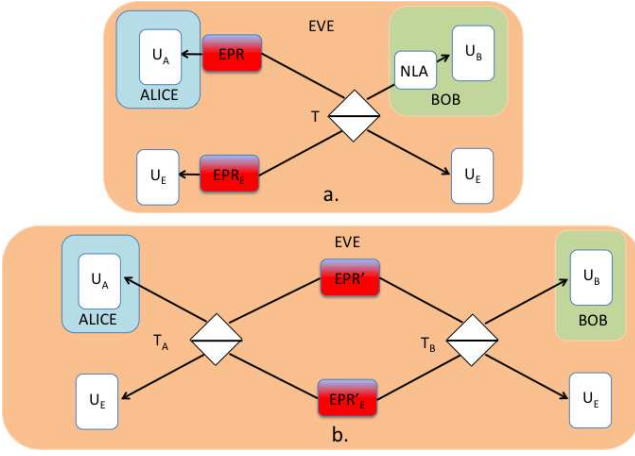}
\caption{\label{equiv} Equivalent eavesdropper attack in the presence on an
  NLA: a) In the original scenario Alice and Eve each create EPR which they
  mix on a beamsplitter of transmission $T$ with the transmitted mode being
  sent to Bob, who uses an NLA. b) The correlations generated by this are
  identical to an equivalent scenario where Eve mixes her unused EPR arm with
  Alice's mode on a beamsplitter of transmission $T_A$ while the original
  beamsplitter changes transmission to a different value $T_B$. Both effective
  EPR pairs also increase in entanglement.}
\end{figure}
We consider the situation where Eve makes an entangling cloner attack as per
the upper panel of Fig.\ref{equiv}, mixing one arm of her own EPR with Bob's
mode on a beamsplitter of transmission $T$.  Bob subsequently applies an NLA
before detection.  The initial 4-mode covariance matrix looks like,
\begin{equation}
{\bf \Sigma} = \matfour{V\hspace{2mm}\mathbb{I}_2&c_{AB}\hspace{2mm}\sigma_z&0&0}{c_{AB}\hspace{2mm}\sigma_z&V\hspace{2mm}\mathbb{I}_2&0&0}{0&0&V_E\hspace{2mm}\mathbb{I}_2&c_{E1E2}\hspace{2mm}\sigma_z}{0&0&c_{E1E2}\hspace{2mm}\sigma_z&V_E\hs  \mathbb{I}_2}
\end{equation}
with $c_{AB} = \sqrt{V^2-1}$ and $c_{E1E2} = \sqrt{V_E^2-1}$.  The final CM is
obtained by enacting a beamsplitter transformation between Bob's mode (B) and
the first eavesdropping mode (E1) that is, ${\bf \Sigma}\rightarrow{\bf
  BS}_{BE1}(T){\bf \Sigma}{\bf BS}_{BE1}^T(T)$ and then substituting this into
Eq.\ref{geq}.  This results in,

\begin{eqnarray}
{\bf \Sigma}_{NLA} =  \matfour{V_A\hspace{2mm}\mathbb{I}_2&c_{AB}\hspace{1mm}\sigma_z&c_{AE1}\hspace{1.0mm}\sigma_z&c_{AE2}\hspace{1mm}\mathbb{I}_2}{c_{AB}\hspace{1mm}\sigma_z&V_B\hspace{2mm}\mathbb{I}_2&c_{BE1}\hspace{1mm}\mathbb{I}_2&c_{BE2}\hspace{2mm}\sigma_z}{c_{AE1}\hspace{1.0mm}\sigma_z&c_{BE1}\hspace{1mm}\mathbb{I}_2&V_{E1}\hspace{2mm}\mathbb{I}_2&c_{E1E2}\hspace{1mm}\sigma_z}{c_{AE2}\hspace{1mm}\mathbb{I}_2&c_{BE2}\hspace{1mm}\sigma_z&c_{E1E2}\hspace{1mm}\sigma_z&V_{E2}\hspace{2mm} \mathbb{I}_2}
\end{eqnarray}
where,
\eqn{V_A \ee V + T + (1-T)V V_E \nn\\
&+& g^2 ( V- T - (1 - T) V V_E)\nn\\
V_B \ee TV+(1 - T)V_E +1 \nn\\
&+& g^2 (TV +(1- T)V_E) - 1) \nn\\
V_{E1} \ee (1-T)V + TV_E +VV_E \nn\\
&+& g^2((1-T)V + TV_E - VV_E)\nn\\
V_{E2} \ee V_E+1 + T(VV_E-1) \nn\\
&+& g^2(V_E - 1 - T(VV_E-1))\nn \\
c_{AB} \ee 2 g \sqrt{T(V^2-1)}\nn\\ 
c_{AE1} \ee -(V_E+1)\sqrt{(1-T((V^2-1)} \nn\\
&+& g^2(V_E - 1 - T(VV_E-1))\nn\\
c_{BE1} \ee 2g\sqrt{(1-T)T}(V-V_E)\nn\\
c_{BE2} \ee 2g\sqrt{(1-T)(V_E^2-1)}\nn\\
c_{E1E2} \ee 2g\sqrt{(1-T)(V_E^2-1)}\nn\\
c_{AE2} \ee (g^2-1)\sqrt{(1-T)T(V^2 - 1)(V_E^2 - 1)} \label{nlacm}}
The elements of the reduced CM shared between Alice and Bob are exactly the
same as in the previous section, however if we consider the correlations with
the eavesdropper we notice an extremely unusual feature. Ordinarily for any
Gaussian channel acting solely upon Bob's side the correlation between Alice's
mode and the eavesdroppers non-interacted mode, $c_{AE2}$, is identically
zero.  However examining this term in Eq.\ref{nlacm} we find that whenever
$g\neq1$ there are correlations despite the fact these modes never interacted.

Such correlations could never be reproduced by an effective setup of the form
of the top panel of Fig.\ref{equiv} so we are motivated to construct an
equivalent setup including an eavesdropping attack on both Alice's and Bob's
modes as per the bottom panel of Fig.\ref{equiv}.  This would produce a CM
given by,
\eqn{{\bf \Sigma}_{\mathrm{equiv}} \ee {\bf BS}_{AE2}(T_A){\bf BS}_{BE1}(T_B) {\bf \Sigma} \nn\\
&&{\bf BS}_{BE1}^T(T_B){\bf BS}_{AE2}^T(T_A)\label{equiv2}}
If we equate Eq.\ref{nlacm} and Eq.\ref{equiv2} some lengthy algebra does
result in a unique solution for the effective scenario where Eve interacts one
EPR mode with Alice and one with Bob.  The expressions for these effective
parameters are given in detail in Appendix A.

\begin{figure}[htbp]
\centering
\includegraphics[width = 8.5cm]{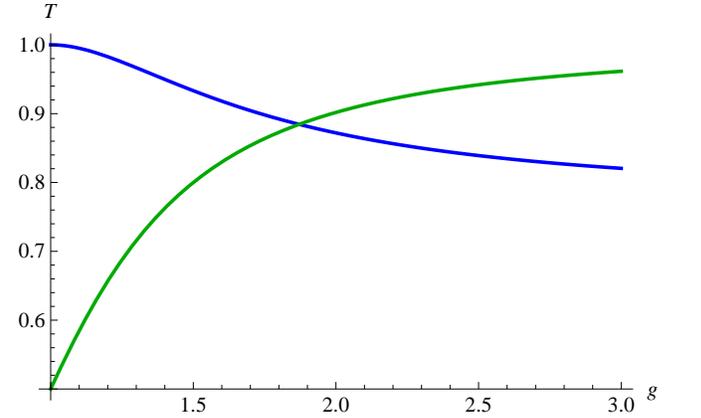}
\caption{\label{tatab} Effective channel transmission Alice's ($T_A$) and
  Bob's ($T_B$) side as a function of the NLA gain. The input parameters are
  $T=0.5$ and $V_E = 1.1$. As the NLA gain increases the channel on Alice's
  side worsens ($T_A$ decreases) whereas the channel on Bob's side improves
  ($T_B$ increases) asymptoting to a perfect channel on Bob's side with the
  attack now exclusively on Alice's side. }
\end{figure}

Considering the same initial state and channel as the previous section the
effective channel parameters reveal an intriguing conclusion about the effect of
the NLA.  As the NLA gain increases the channel on Bob's side improves
(i.e. the effective transmission increases) whereas the converse is true of
Alice's side as shown in Fig.\ref{tatab}.  In fact for these parameters the
attack on Bob's side disappears almost entirely.

\begin{figure}[htbp]
\centering
\includegraphics[width = 8.5cm]{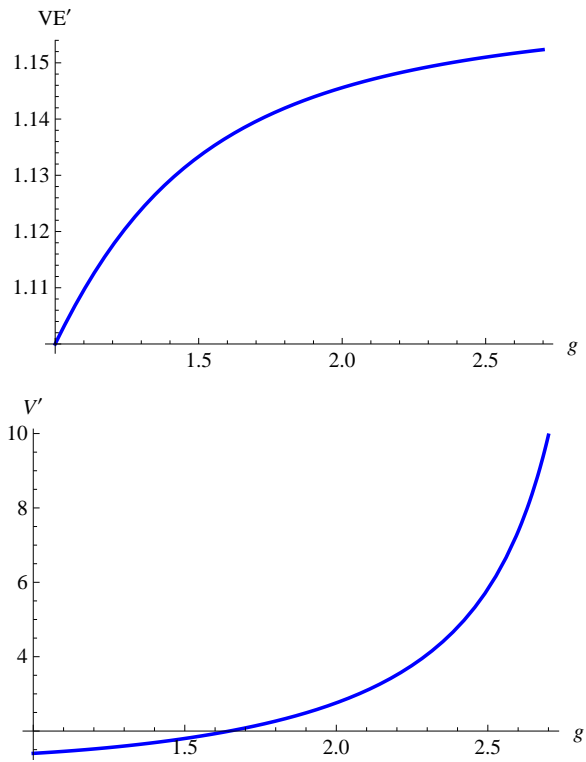}
\caption{\label{VVE} Effective entanglement for Alice (bottom panel) and Eve
  (top panel) paramaterised by their variance as a function of the NLA
  gain. The input parameters are $T=0.5$ and $V_E = 1.1$. The effective
  entanglement of both parties increases as a function of the gain.}
\end{figure}

Finally we also see that both Alice and Eve's effective initial entanglement
is increased though not equitably. The variance characterising Eve's effective
EPR increases modestly as the gain approaches it's maximum whereas Alice's
diverges as expected from the previous thermal state results.

\section{Conclusions \label{c}}
In conclusion we have considered the phase space representation of an ideal
NLA, gaining insight into the regimes resulting in physical outputs for input
states other than coherent states, explaining this in terms of the relative
divergence of the NLA and the target state in phase space.  For Gaussian
states we have derived compact analytic formula's for the action of up to $N$
amplifiers upon $N$-mode states. We have given explicit results for important
examples in one and two modes and analysed the strength and purity of
entanglement of EPR states through general Gaussian decoherence, demonstrating
that our methods allow for the swift identification of the best strategy for a
given protocol. Finally we uncovered some intriguing effects on correlations
between two, two-mode entangled states with the counterintuitive result that
under amplification an interaction is moved from one side to another.

Future work of great practical importance will be to investigate various
proposals for implementations of the NLA, especially the promising
post-selection approaches of \cite{Walk:2012p5481,Fiurasek:2012p5315}, to
determine corresponding success probabilities which are critical for protocols
where the rate is of great importance such as QKD and metrology. Finally other
results enabled by this work will include revisiting various canonical quantum
information tasks such as cloning and state discrimination.

%\ack
{\it Acknowledgements.---} 
The authors would like to thank S. Rahimi-Keshari for useful discussions.
This research was conducted by the Australian Research Council Centre of
Excellence for Quantum Computation and Communication Technology (Project
number CE11000102)

\begin{appendix}
\section{4-mode equivalent circuit}
Here we derive in detail the parameters for the effective circuit shown in the
bottom panel of Fig.\ref{equiv}.  We start with two EPR pairs belonging to Eve
and Alice, paramaterised by variances $V',V_E'$ respectively. To calculate the
necessary output CM's we need the NLA transform given by Eq.\ref{geq} and the
4-mode version of the beamsplitter transform.  This induces correlations
between the two target modes and acts as the identity upon the others.  For
example the 4-mode beamsplitter matrix acting upon modes 2 and 3 with
transmission $T$ is,
\eq{{\bf BS}_{2,3}(T) =  \matfour{\hspace{2mm}\mathbb{I}_2&0&0&0}{0&\sqrt{T}\hspace{2mm}\mathbb{I}_2&-\sqrt{1-T}&0}{0&\sqrt{1-T}\hspace{2mm}\sigma_z&\sqrt{T}\hspace{2mm}\mathbb{I}_2&0}{0&0&0&\mathbb{I}_2} \label{bs}}
Note that the minus sign that appears on one of the correlation terms is
essentially a choice of convention, or corresponds to a choice to swap which
mode enters which port of the beamsplitter.

In this purified version of the protocol the transmission through the channel
simply corresponds to mixing modes B (2) and and E1(3) via \ref{bs}
\eq{{\bf \Sigma}_{\mathrm{ch}} = {\bf BS}_{B,E1}(T) {\bf \Sigma} {\bf BS}_{B,E1}^T(T)}
and the substituting this into Eq.\ref{geq} which directly gives the terms in
Eq.\ref{nlacm}.  We now calculate the equivalent circuit CM given by Eq.\ref{equiv2}
which has the block diagonal form of Eq.\ref{cm} with

\begin{widetext}
\eqn{V_A \ee V_E' + T_A (V'-V_E')\nn\\
V_B \ee V_E' + T_B  (V'-V_E')\nn\\
V_{E1} \ee T_B (V_E' - V') + V'\nn \\
V_{E2} \ee T_A (V_E' - V') + V'\nn\\
c_{AB} \ee -\sqrt{(1 - T_A) (1 - T_B) (V_E'^2-1)} + \sqrt{T_A T_B (V'^2-1 )}\nn\\
c_{AE1} \ee -\sqrt{(1 - T_A) T_B (V_E'^2-1)} - \sqrt{T_A (1 - T_B) (V'^2-1)}\nn\\
c_{BE1} \ee \sqrt{(1 - T_B) T_B} (V_E' - V')\nn\\
c_{BE2} \ee \sqrt{T_A (1 - T_B) (V_E'^2-1)} + \sqrt{(-1 + T_A) T_B (1 - V'^2)}\nn\\ c_{E1E2} \ee\sqrt{T_A T_B (V_E'^2-1)} - \sqrt{(1 - T_A) (1 - T_B) (V'^2-1)}\nn\\
c_{AE2} \ee \sqrt{(1 - T_A) T_A} (V'-V_E') \label{abcequiv}}
\end{widetext}
We now wish to set this expression equal to Eq.\ref{nlacm} simultaneously
solve for $V',V_E',T_A,T_B$.  In practice one an proceed by considering only a
few terms and then checking that the solution satisfies all terms.  By
equating the variance terms one can swiftly solve for three of the parameters
in terms of the fourth and the input variables obtaining,

\begin{widetext}
\eq{\left(\begin{array}{c} 
T_A\\
\\
T_B\\
\\
V_E'
\end{array}
\right )
= \left(
\begin{array}{c}
 \frac{(1+V_E) (-1+V')+T (1-V V_E+V V'-V_E V')+g^2 ((1-V_E) (1+V')+T (-1+V V_E-V V'+V_E V'))}{(-1-V) (1+V_E)+2 (1+T (V-V_E)+V_E) V'+g^2 ((-1+V) (-1+V_E)-2 (-1+T (V-V_E)+V_E) V')} \\
 \\
 \frac{\left(-1-g^2\right) T V_E+\left(1+g^2+\left(-1+g^2\right) (-1+T) V_E\right) V'+V \left(-1+T-V_E+T V'+g^2 (-1+T+V_E-T V')\right)}{(-1-V) (1+V_E)+2 (1+T (V-V_E)+V_E) V'+g^2 ((-1+V) (-1+V_E)-2 (-1+T (V-V_E)+V_E) V')} \\
 \\
 \frac{(-1-V) (1+V_E)+(1+T (V-V_E)+V_E) V'+g^2 (1+V'+V_E (-1+(-1+T) V')+V (-1+V_E-T V'))}{-1-T V+(-1+T) V_E+g^2 (-1+T (V-V_E)+V_E)}
\end{array}
\right)}
\end{widetext}
Considering, say, the $c_{AE2}$ correlation term we find a quadratic with the
two solutions corresponding to a permutation of the roles of Alice and Eve.
Considering any of the other correlation terms will yield only one consistent
solution.  One further subtlety is apparent upon considering the $c_{AE2}$
term.  In particular the expression Eq.\ref{nlacm} is always positive whereas
the corresponding term in Eq.\ref{abcequiv} changes sign depending upon the
relative magnitude of $V$ and $V_E$.  The resolution to this is that for
situations where the original parameters are such that $V_E>V$ we must change
the phase convention for the beamsplitter on Alice's side.  This corresponds
to moving the minus sign in Eq.\ref{bs} to the other correlation term.  The
two solutions for the remaining parameter $V'$ are,
\begin{widetext}
\eqn{V' = \left\{ 
\begin{array}{c}
\frac{g^2 (-1+V) (-1+V_E)-(1+V) (1+V_E)-\sqrt{B^2+4 C}}{2 \left(-1-T V+(-1+T) V_E+g^2 (-1+T (V-V_E)+V_E)\right)}, \hs V>V_E\\
\\
\frac{g^2 (-1+V) (-1+V_E)-(1+V) (1+V_E)+\sqrt{B^2+4 C}}{2 \left(-1-T V+(-1+T) V_E+g^2 (-1+T (V-V_E)+V_E)\right)}, \hs V<V_E
\end{array}
\right .}
where
\eqn{B \ee\nn 1+V+V_E+V V_E+g^2 (-1+V+V_E-V V_E)\\
C \ee \nn \left(-1-T V+(-1+T) V_E+g^2 (-1+T (V-V_E)+V_E)\right)\left(\left(1-g^2\right) T V_E+V \left(1-T+V_E+g^2 (-1+T+V_E)\right)\right)}
\end{widetext}

Direct substitution then confirms that for all input parameters we have a
unique set of parameters that yield an identical CM to that created by the
NLA.
\end{appendix}
\end{document}